\documentclass[12pt, onecolumn]{IEEEtran}
\usepackage{epsfig,graphics}

\begin{document}
\title{\Large\bfseries Joint source and channel coding for MIMO systems: \\ Is it better
to be robust or quick? \thanks{The first author is with Goldman Sachs, the second author is with Stanford University, and the third author is with Princeton University. This research was supported by the Office of Naval Research under Grant N00014-05-1-0168, by DARPA's ITMANET program under Grant 1105741-1-TFIND, and by the National Science
Foundation under Grants ANI-03-38807 and CNS-06-25637.}}
\author{Tim Holliday, Andrea J. Goldsmith, and H. Vincent Poor}
\maketitle
\thispagestyle{empty}
\pagestyle{empty}

\vspace{-.2in}

\begin{abstract}
We develop a framework to optimize the tradeoff between
diversity, multiplexing, and delay in MIMO systems to 
minimize end-to-end distortion. The goal is to find the optimal balance
between the increased data rate provided by antenna multiplexing,
the reduction in transmission errors provided by antenna diversity and automatic repeat request (ARQ),
and the delay introduced by ARQ. 
We first focus on the diversity-multiplexing tradeoff in MIMO systems, and develop analytical
results to minimize distortion of  a vector
quantizer concatenated with a space-time MIMO channel code. In the high SNR regime we obtain a closed-form expression for the end-to-end distortion as a function
of the optimal point on the diversity-multiplexing tradeoff curve.
For large but finite SNR we find this optimal point via convex optimization.  
The same general framework can also be used
to minimize end-to-end distortion for a broad class of
practical source and channel codes, which we illustrate with an example.  

We then consider MIMO systems using ARQ retransmission to
provide additional diversity at the expense of delay. 
We show that for
sources without a delay constraint, distortion is minimized
by maximizing the ARQ window size.  This results in an ARQ-enhanced
multiplexing-diversity tradeoff region, with distortion minimized over this region 
in the same manner as without ARQ. 
However, under a source delay constraint the problem formulation changes to
account for delay distortion associated with
random
message arrival and random ARQ completion times.  
Moreover, the simplifications associated with a high SNR assumption break down for this analysis, 
since retransmissions, and the delay they cause, become rare events. 
We thus use a dynamic programming
formulation to capture the channel diversity-multiplexing tradeoff at finite
SNR as well as the random arrival and retransmission dynamics. This fomulation
is used to solve for the optimal multiplexing-diversity-delay tradeoff to minimize
end-to-end distortion associated with the source encoder, channel, and ARQ retransmissions. Our results show that a delay-sensitive system
should adapt its operating point on the diversity-multiplexing-delay tradeoff region to
 the system dynamics. We provide numerical results
that demonstrate significant performance gains of this adaptive policy
over a static allocation
of diversity/multiplexing in the channel code and a static ARQ window size. 

\vspace{.08in}
\end{abstract}

\section{Introduction}

Multiple antennas can significantly improve the performance of
wireless systems. In particular, with channel knowledge at the
receiver a data rate increase equal to the minimum number of
transmit/receive antennas can be obtained by multiplexing data
streams across the parallel channels associated with the channel
gain matrix. Alternatively, multiple antennas enable 
transmit and/or receive diversity which decreases the probability of
error. In a landmark result Zheng and Tse \cite{Zheng} developed a
rigorous fundamental tradeoff between the data rate increase possible via multiplexing versus
the channel error probability reduction possible via diversity,
characterizing how a higher spatial multiplexing gain leads to
lower diversity and vice versa. The main result in \cite{Zheng} is
an explicit characterization of the diversity-multiplexing
tradeoff region. This result generated much activity in finding
diversity-multiplexing tradeoffs for other channel models as well
as design of space-time codes that achieve any point on the
tradeoff region \cite{Azarian,ElGamal,ElGamal2,Kuhn,Lu,Wornell}.
The diversity-multiplexing tradeoff was also extended to the multiple access channel
in \cite{Tse}.  Delay provides a third dimension in the tradeoff region, and this
dimension was explored for MIMO channels based on the automatic repeat request (ARQ)
protocol in \cite{ARQ}.  In particular, this work characterized the three-dimensional
tradeoff between diversity, multiplexing, and ARQ-delay for MIMO systems.

Our goal in this paper is to answer the following question:
``Given the diversity-multiplexing-delay tradeoff region, where should a system
operate on this region?''.  In order to answer this question we require a
performance metric from a layer above the physical layer; while physical layer
tradeoffs are based on the channel model, the optimization between these tradeoffs
depends on what is most important for the application's end-to-end performance.
The higher layer metric of interest in
this paper will be end-to-end distortion.  Specifically, our
system model consists of a lossy source encoder concatenated with
a MIMO channel encoder and, in the last section, an ARQ retransmission
protocol.  Our goal is to determine the optimal
point on the diversity-multiplexing or diversity-multiplexing-delay
tradeoff region that minimizes the combined
distortion due to the source compression, channel, and delays in the end-to-end system. 

Our problem formulation differs from the 
Shannon-theoretic joint source-channel coding problem  in that
we do not assume asymptotically long block lengths for either
the source or channel code. 
In particular, the traditional joint source/channel code formulation 
assumes stationary and ergodic
sources and channels in the asymptotic regime of large source dimension
and channel code blocklength. Shannon showed that under these assumptions
the source should be encoded at a rate just below
channel capacity and then transmitted over the channel at this rate.  
Since the rate is less than capacity, the channel introduces negligible
error, hence the end-to-end distortion equals the distortion introduced by
compressing the source to a rate below the channel capacity. 
Shannon's well-known separation theorem indicates that this transmission scheme is optimal for minimizing end-to-end distortion
and does not require any coordination
between the source and channel coders or decoders other than agreeing on the channel
transmission rate \cite{Cover,Csiszar}.

Our joint source/channel code formulation is fundamentally different 
from Shannon's since we
 assume a finite blocklength for the channel code.
 This assumption is inherent to the diversity-multiplexing tradeoff since, without
finite blocklength, the channel introduces negligible error and hence the diversity gain
in terms of channel error probability is meaningless. 
The finite blocklength
guarantees there is a nonnegligible probability of error in the
channel transmission.  Thus there is a tradeoff between resolution at the source encoder and
robustness at the channel encoder: limiting source distortion requires
a high-rate source code, for which the multiple antennas of the channel must be
used mainly for multiplexing. 
Alternatively, the source can be encoded at a lower rate with more distortion, and then
the channel error probability can be reduced through increased diversity. Our joint
source/channel code must determine the best tradeoff between these two to minimize
end-to-end distortion. When retransmission is
possible and the source is delay-sensitive, there is an additional tradeoff between reducing channel errors through retransmissions versus the
delay these retransmissions entail. 

Joint source/channel code optimization 
for the binary symmetric channel (BSC) 
with finite blocklength channel codes and asymptotically high
source dimension
was previously studied in \cite{Hochwald}. We will use several key ideas and results from
this prior work in our asymptotic analysis, in particular its decomposition of 
 end-to-end distortion 
into separate components associated with either the source code
or the channel code. By applying this decomposition to MIMO channels instead of the BSC,
 we obtain 
 the optimal operating point on the Zheng/Tse diversity-multiplexing tradeoff region in the 
asymptotic limit
of high source dimension and channel SNR. For any SNR the MIMO channel under multiplexing
can be viewed as a parallel channel, and
source/channel coding for parallel channels has been previously explored in
\cite{Laneman}.  That work differs from ours in that the source models 
were not high dimensional and the nonergodic parallel channels 
 did not have the same diversity-multiplexing tradeoff
characterization as in a MIMO system.

We first develop a closed-form expression for the optimal
``distortion exponent'', introduced in \cite{Laneman}, under
asymptotically high SNR. Specifically, for a multiplexing rate $r$ and
average distortion measure $D(r)$ we compute
\begin{equation}
d_D^*=\min_{r}\left[\lim_{\mathrm{SNR}\rightarrow\infty}\frac{\log
\bar{D}(r)}{\log \mathrm{SNR}}\right],
\end{equation}
where $d_D^*$ is the optimal exponential rate at which the
distortion goes to zero with SNR. We show that the optimal distortion exponent
corresponds to a particular point on the diversity-multiplexing
tradeoff curve that is determined by the source characteristics.  
We also demonstrate there is no loss in optimality for separate source and channel
encoding and decoding given the channel transmission rate. Our optimization framework can
also be used to optimize the diversity-multiplexing tradeoff at finite SNR, however the
solution is no longer in closed-form and must be found
using tools from convex optimization. We extend this general optimization framework
to a wide variety of practical source-channel codes in 
non-asymptotic regimes. 

We next consider the impact of ARQ retransmissions and their associated delay.
When the source does not have a delay constraint, the ARQ delay incurs
 no cost in terms of additional distortion. Hence, the ARQ protocol
should use the maximum window size to enhance the diversity-multiplexing
tradeoff region associated with the MIMO channel alone. The large window
size essentially allows coding over larger blocklengths than without ARQ,
which from Shannon theory does not reduce data rate, only probability of error. In the high
SNR regime the optimal distortion exponent for the diversity-multiplexing
tradeoff region enhanced by ARQ is
found in the same manner as without ARQ. 
Not surprisingly, a delay constraint on the source changes the problem
considerably, since the source burstiness and queuing delay must now
be incorporated into the problem formulation. These characteristics
are known to be a significant obstacle in merging analysis of 
the fundamental limits at the physical
layer with end-to-end
network performance \cite{Ephremides}. 
In this setting the simplicity associated with the 
high SNR regime breaks down, since at high SNR retransmissions and their associated delay 
have very low probability, which 
essentially removes the third dimension of delay in our tradeoff region. 
We thus use dynamic programming to model and optimize over the system dynamics as well
as the fundamental physical layer tradeoffs
to minimize end-to-end distortion of a MIMO channel with ARQ. 

The remainder of this paper is organized as follows. In the next
section we present the channel model and summarize the
diversity-multiplexing tradeoff results from \cite{Zheng}.  In
Section III we develop our source encoding framework and apply the
MIMO channel error probability results of \cite{Zheng} to the
upper and lower bounds on end-to-end distortion 
of \cite{Hochwald}.  Section IV
obtains a closed-form expression for the optimal
operating point on the MIMO channel diversity-multiplexing tradeoff curve in the high
SNR regime to minimize end-to-end distortion. This optimal point is also found
for large, but finite, SNR using convex optimization. 
In Section V we present a similar formulation for
optimizing diversity and multiplexing in progressive video
transmission using space-time codes.  ARQ retransmission and its corresponding
delay is considered in Section VI, where a dynamic programming formulation
is used to optimize the operating point on the diversity-multiplexing-delay
tradeoff region for minimum end-to-end distortion of delay-constrained sources.  
A summary and closing thoughts are provided
in Section VII.

\section{Channel Model}

We will use the same channel model and notation as in \cite{Zheng}.
 Consider a wireless channel with $M$ transmit antennas and $N$
receive antennas. The fading coefficients $h_{ij}$ that model the
gain from transmit antenna $i$ to receive antenna $j$ are
independent and identically distributed (i.i.d.) complex Gaussian
with unit variance. The channel gain matrix $\mathbf{H}$ with
elements $H(i,j)=(h_{ij}:i\in\{1,\ldots M\},j\in\{1,\ldots,N\})$
is assumed to be known at the receiver and unknown at the
transmitter.  We assume that the channel remains constant over a
block of $T$ symbols, while each block is i.i.d. Therefore, in
each block we can represent the channel as
\begin{equation} \label{channel}
\mathbf{Y}=\sqrt{\frac{\mathrm{SNR}}{M}}\mathbf{HX}+\mathbf{W}, 
\end{equation}
where $\mathbf{X}\in\mathcal{C}^{M\mathrm{x}T}$ and
$\mathbf{Y}\in\mathcal{C}^{N\mathrm{x}T}$ are the transmitted and
received signal vectors, respectively.  The additive noise vector
$\mathbf{W}$ is i.i.d. complex Gaussian with unit variance.

We construct a family of codes for this channel
$\{C(\mathrm{SNR})\}$ of block length $T$ for each $\mathrm{SNR}$
level. Define $P_e(\mathrm{SNR})$ as the average probability of
error and $R(\mathrm{SNR})$ as the number of bits per symbol for
the codebook.  A channel code scheme $\{C(\mathrm{SNR})\}$ is said
to achieve multiplexing gain $r$ and diversity gain $d$ if
\begin{equation}
\lim_{
\mathrm{SNR}\rightarrow\infty}\frac{R(\mathrm{SNR})}{\log_2
\mathrm{SNR}}=r,
\end{equation}
and
\begin{equation}
\lim_{ \mathrm{SNR}\rightarrow\infty}\frac{\log_2
P_e(\mathrm{SNR})}{\log_2 \mathrm{SNR}}= -d.
\end{equation}
All logarithms we consider will have base 2 and we therefore suppress
this base notation in the remainder of the paper.  For each $r$
we define the optimal diversity gain $d^*(r)$ as the supremum of
the diversity gain achieved by any scheme. The main result from
\cite{Zheng} that we will use in the next section is
summarized in the following statement.

\bigskip
\noindent\textbf{Diversity-Multiplexing Tradeoff \cite{Zheng}:}
Assume the block length satisfies $T\ge M+N-1$. Then the optimal
tradeoff between diversity gain and multiplexing gain is the
piecewise-linear function connecting the points
$d^*(r)=(M-r)(N-r)$, for integer values of $r$ such that $0\le
r\le \min(M,N)$.  This function $d^*(r)$ is plotted in Figure 1.

\bigskip
In the Zheng/Tse framework the rate of the codebook $\{C(\mathrm{SNR})\}$
must scale with $\log \mathrm{SNR}$, otherwise the multiplexing
gain will go to zero. Hence, in the following sections we will
assume, without loss of generality, that the rate of the codebook
is $Tr\log \mathrm{SNR}$ for any choice of $0\le r\le\min(M,N)$
and block length $T$.  We also assume that the codebook achieves
the optimal diversity gain $d^*(r)$ for any choice of $r$.  
Codes achieving the optimal diversity-multiplexing tradeoff for MIMO
channels have been investigated in many works, including
\cite{ElGamal2,ElGamal,ElgamalDamen,Sethur} and the references therein.

\begin{figure}
\begin{center}
\psfig{figure=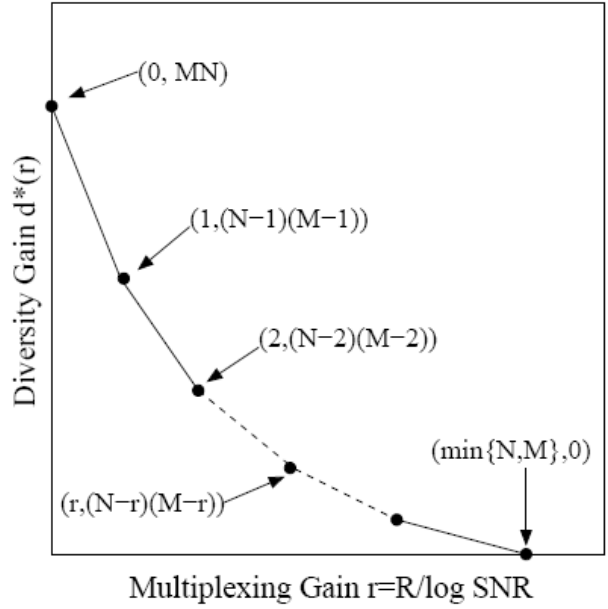, angle=0, width=3.5in}
\end{center}
\caption{ The optimal diversity-multiplexing tradeoff for $T\ge
M+N-1$.}
\end{figure}

\section{End-to-End Distortion}

This section presents our system model for the end-to-end
transmission of source data.  We use the same source coding model
as \cite{Hochwald} in order to exploit their decomposition of 
 end-to-end distortion 
into separate source and channel distortion components.  We assume the original source
data $u$ is a random variable with probability density $h(u)$,
which has support on a closed and bounded subset of $\Re^k$ with
non-empty interior.  An $s$-bit quantizer is applied to $u$ via
the following transformation:
\begin{equation}
Q(u)=\sum_{i=1}^{2^s}v_i I_{[A_i]}(u),
\end{equation}
where $I_{[A_i]}(u)=I[u \in A_i]$ is the standard indicator function, and
$\{A_i\}_{i=1}^{2^s}$ is a partition of $\Re^k$ into disjoint
regions.  Each region $A_i$ is represented by a single codevector
$v_i$.  The $p$th-order distortion due to the encoding process is
\begin{equation}
D_s(Q)=\sum_{i=1}^{2^s}\int_{A_i}||u-v_i||^p h(u)du,
\end{equation}
where $||u-v_i||^p$ is the $p$th power of the Euclidian norm.

We assume that the rate of the channel codebook
$\mathcal{C}\{\mathrm{SNR}\}$ is matched to the rate of the
quantizer (i.e. $s=Tr\log \mathrm{SNR}$). Each codevector from the
quantizer $v_1,\ldots,v_{_{2^s}}$ is mapped into a codeword from
$\mathcal{C}\{\mathrm{SNR}\}$ through a permutation mapping $\pi$.
We assume the mapping $\pi$ is chosen equally likely at random
from the $2^s!$ possibilities.  The codeword $\pi(i)$ is
transmitted over the channel described in Section 2 and decoded at
the receiver. Let $q(\pi(j)|\pi(i))$ be the probability that
codeword $\pi(j)$ is decoded at the receiver given that $\pi(i)$
was transmitted. The probability $q(\cdot|\cdot)$ will depend on
the $\mathrm{SNR}$, the quantizer $Q$'s codeword set, and the permutation mapping
$\pi$. Hence, we can write the total end-to-end distortion as
follows:
\begin{equation}\label{DT}
D_{\tau}(Q,\mathrm{SNR},\pi)=\sum_{i=1}^{2^s}\sum_{j=1}^{2^s}q(\pi(j)|\pi(i))\int_{A_j}||u-v_j||^p
h(u)du.
\end{equation}

Ideally we would like to be able to analyze the distortion
averaged over all index assignments and possibly remove the
dependence on $h$ and $Q$.  In general we cannot find a closed
form expression for this distortion due to the dependence on
$Q$'s codewords, $\pi$, $h$, and the SNR. However, given our matched source and
channel rate $s=Tr\log \mathrm{SNR}$, is clear that we have a
tradeoff between transmitting at a high data rate to reduce source 
distortion and transmitting at a low data rate to reduce channel errors.
In particular, if we run full multiplexing in the MIMO channel (i.e. set
$r=\min(M,N)$) we can use a large $s$.  This would result in low
distortion at the source encoder but possibly create many
transmission errors. Conversely, we could use full diversity in
the channel (i.e. set $d=MN$) to combat errors and then suffer the
distortion from a low value of $s$.  Between the two extremes lies
a source code rate $s$ and a corresponding channel multiplexing rate $r$ that
minimizes (\ref{DT}).

Although we cannot find a simple general expression for
$D_{\tau}(Q,\mathrm{SNR},\pi)$, in the following subsections we
will determine tight asymptotic bounds for the distortion through
the use of high-resolution source coding theory and high-SNR
analysis of the MIMO channel. In addition, as the $\mathrm{SNR}$
approaches infinity we will find a simple expression for the
optimal choice of $r$ and $s$ that depends only on the block
length $T$, source dimension $k$, number of transmit antennas $M$,
and number of receive antennas $N$.

The high-resolution asymptotic regime is often used in source
coding theory to obtain analytical results, since the performance
characteristics of many encoder types are well understood in this
regime \cite{Zeger}.  Moreover, it has been show that the high
resolution asymptotics often provide a good approximation for
non-asymptotic performance \cite{Max,Trushkin}.  As described in
\cite{Zeger}, we say that a quantizer $Q$ operates in the
high-resolution asymptotic regime if its noiseless distortion
asymptotically approaches
\begin{equation}\label{asym}
D_s(Q)=2^{-ps/k+O(1)},
\end{equation}
as $s$ goes to infinity, where the $O(1)$ term in (\ref{asym}) may
depend on $p$, $k$, and $s$.  Many practical quantizers achieve this
asymptotic distortion, e.g. uniform and lattice-based quantizers
\cite{Bucklew,Zador}.
This high-resolution asymptotic regime is quite accurate for our system model
since we require the rate of our channel codebook
$\{C(\mathrm{SNR})\}$ to scale as $r\log \mathrm{SNR}$. Hence, at asymptotically high
SNR, the
source coder will receive an increasing number of bits, thereby approaching its
high-resolution regime.

In the next two subsections we will construct upper and lower asymptotic bounds
for the end-to-end average distortion of our system.  The starting point for both 
bounds comes from the analysis of \cite{Hochwald}. In Section IV
we will show that these bounds are tight and find the optimal
multiplexing rate that minimizes distortion in the high SNR
regime.

\subsection{Upper Bound for Distortion}

We first construct an upper bound for the end-to-end distortion
(\ref{DT}) that depends on $\pi$.  As shown in \cite{Hochwald},
\begin{eqnarray}
D_{\tau}(Q,\mathrm{SNR},\pi)&=&\sum_{i=1}^{2^s}\sum_{j=1}^{2^s}q(\pi(j)|\pi(i))\int_{A_i}||u-v_j||^p
h(u)du \nonumber \\
&=&\sum_{i=1}^{2^s}q(\pi(i)|\pi(i))\int_{A_i}||u-v_j||^p h(u)du \nonumber \\
&+& \sum_{j,i=1, i\ne
j}^{2^s}q(\pi(j)|\pi(i))\int_{A_i}||u-v_j||^p h(u)du \nonumber \\
&\le&\sum_{i=1}^{2^s}\int_{A_i}||u-v_j||^p
h(u)du+O(1)\sum_{i=1}^{2^s}P(A_i)\sum_{j=1,j\ne
i}^{2^s}q(\pi(j)|\pi(i)) \nonumber \\
&\le& D_s(Q)+O(1)\max_{i}P_{e|\pi(i)}(\mathrm{SNR}),\label{UB}
\end{eqnarray}
where $P_{e|\pi(i)}$ is the probability of codeword error given
that codeword $\pi(i)$ was transmitted.  This bound essentially
splits (\ref{DT}) into two pieces; one corresponding to correctly
received channel codewords and the other corresponding to
erroneous channel decoding.  The term corresponding to correct
transmission is bounded by the noiseless distortion $D_s(Q)$ while
the term corresponding to errors is bounded by a
constant\footnote{This term is $O(1)$ because our source is
bounded.} multiplied by the channel codeword error probability.

By construction, the rate of our channel codebook (and hence the
source encoder) is $s=Tr\log \mathrm{SNR}$, therefore
\begin{equation} \label{DsQ} 
D_s(Q)=2^{-ps/k+O(1)}=2^{-\frac{pTr}{k}\log \mathrm{SNR}+O(1)}
\end{equation}
as $s$ approaches infinity or, equivalently, as $\log \mathrm{SNR}$ approaches
infinity. 
 In order to bound the probability of codeword
error we need a few quantities from \cite{Zheng}.  For the channel
defined in (\ref{channel}), let $P_{out}(r\log\mathrm{SNR})$ and
$d_{out}(r)$ be the outage probability and outage exponent that
satisfy
\begin{equation}
P_{out}(r\log \mathrm{SNR})=2^{-d_{out}(r)\log\mathrm{SNR}+o(\log
\mathrm{SNR})} \label{Pout}.
\end{equation}
The exponent $d_{out}(r)$ can be directly computed and the
equation for doing so is presented in \cite{Zheng}.

We can also bound the probability of error with no outage through
\begin{equation}
P(\mathrm{error}, \;\mathrm{no}\;\mathrm{ outage})\le
2^{-d_{G}(r)\log\mathrm{SNR}+o(\log \mathrm{SNR})}, \label{Pe}
\end{equation}
where $d_G(r)$ is the exponent associated with choosing the
channel codewords to be i.i.d. Gaussian.  Again, the formula for
computing $d_G(r)$ can be found in \cite{Zheng}.  Then we can
bound the overall probability of error $P_e(\mathrm{SNR})$ by
\begin{eqnarray}
P_e(\mathrm{SNR})&\le&
P_{out}(r\log\mathrm{SNR})+P(\mathrm{error},
\;\mathrm{no}\;\mathrm{
outage}) \nonumber \\
&\le& 2^{-d_{out}(r)\log\mathrm{SNR}+o(\log
\mathrm{SNR})}+2^{-d_{G}(r)\log\mathrm{SNR}+o(\log \mathrm{SNR})}.
\label{Petot}
\end{eqnarray}
With the bound (\ref{Petot}) in hand we may now upper bound the total distortion
by
\begin{equation}\label{DTUB}
D_{\tau}(Q,\mathrm{SNR},\pi)\le 2^{-\frac{pTr}{k}\log
\mathrm{SNR}+O(1)}+O(1)2^{-d_{out}(r)\log\mathrm{SNR}+o(\log
\mathrm{SNR})}+O(1)2^{-d_{G}(r)\log\mathrm{SNR}+o(\log \mathrm{SNR})}.
\end{equation}

Note that the distortion upper bound in
(\ref{DTUB}) does not depend on the source-to-channel
codeword mapping $\pi$, since the bounds (\ref{Pout}) and (\ref{Pe}) as
well as the source distortion (\ref{DsQ}) 
do not depend on this mapping. 
Hence, the bound (\ref{DTUB}) holds for the distortion averaged
over all possible source-codeword mappings, and only depends on the quantizer $Q$ through
the parameters $p$, $s$, and $k$. Thus, by averaging over all source-channel codeword
mappings we get that for any quantizer
$Q$ satisfying (\ref{asym}) in the high resolution asymptotic regime, the
end-to-end average distortion is bounded above by
\begin{eqnarray}
\bar{D}_{\tau}(\mathrm{SNR})&=& E_{\pi}[D_{\tau}(\mathrm{SNR},\pi)]\nonumber \\
&\le &  2^{-\frac{pTr}{k}\log
\mathrm{SNR}+O(1)}+O(1)2^{-d_{out}(r)\log\mathrm{SNR}+o(\log
\mathrm{SNR})}+O(1)2^{-d_{G}(r)\log\mathrm{SNR}+o(\log \mathrm{SNR})}.
\label{DTUBave}
\end{eqnarray}

\subsection{Lower Bound for Distortion}

Our lower bound for distortion will also make use
of a result from \cite{Hochwald}.  Let
$\bar{D}_{\tau}(Q,\mathrm{SNR})$ be the distortion averaged over
all $2^s!$ possible mappings $\pi$.  Then from \cite{Hochwald}
we have
\begin{equation}
\bar{D}_{\tau}(Q,\mathrm{SNR})\ge
2^{-ps/k+O(1)}+O(1)P_e(\mathrm{SNR}).
\end{equation}
Note that as in the upper bound, for any quantizer $Q$ satisfying (\ref{asym}) in
the asymptotic regime, the
lower bound depends on $Q$ only through the parameters $p$, $s$ and $k$. 
However, a key difference between this bound and the upper bound (\ref{DTUB}) is
that it is based on averaging distortion over all source-codeword mappings $\pi$.  In
particular, this bound is based on the assumption that each source-to-channel codeword mapping is random and equally probable (i.e. the probability
of mapping a given
 source codeword to a given channel codeword is uniform). From \cite{Zheng} we may lower bound the error
probability $P_e(\mathrm{SNR})$ via the outage exponent as
\begin{equation}
P_e(\mathrm{SNR})\ge
2^{-d_{out}\log\mathrm{SNR}+o(\log\mathrm{SNR})}.
\end{equation}
Thus our lower bound for average distortion  for any quantizer
$Q$ satisfying (\ref{asym}) in the asymptotic regime of high resolution becomes
\begin{equation}\label{DTLB}
\bar{D}_{\tau}(\mathrm{SNR})\ge
2^{-ps/k+O(1)}+O(1)2^{-d_{out}\log\mathrm{SNR}+o(\log\mathrm{SNR})}.
\end{equation}

\section{Minimizing Total Distortion}

In this section we will optimize the bounds presented in the previous
section and show that they are tight.  In order to achieve
analytical results for the minimum distortion bound we consider
the asymptotic regime of ${SNR}$ approaching infinity.
In general, our total distortion is an exponential sum of the form
\begin{equation} 2^{f(r)\log \mathrm{SNR}}+2^{g(r)\log \mathrm{SNR}}
\label{expsum}, \end{equation} where
we define $f(r)$ as the {\em source distortion exponent} and
$g(r)$ as the {\em channel distortion exponent}. 
We minimize
total distortion in the form of (\ref{expsum})
 by choosing
the exponents $f(r)$ and $g(r)$ to be within $o(1)$ of each other.
The function $f(r)$ depends on the source distortion while
$g(r)$ depends on the channel error probability.
For example, in (\ref{DTLB}), if we assume the bound is tight and 
neglect terms that become negligible at high
SNR, then $f(r)=-pTr/k$ (since $s=Tr \log \mathrm{SNR}$) and
$g(r)= -d_{out}(r)$. 
Note that if the exponents in (\ref{expsum}) are not of the same order then one
term in the sum dominates the other as $ \mathrm{SNR}$
approaches infinity. As we shall see, the fact that these two terms are of the
same order is the key to obtaining a closed-form expression
for the optimal diversity-multiplexing tradeoff point. 

\subsection{Asymptotic Regime}

We first consider the upper bound for total distortion
(\ref{DTUB}).   We need to match the exponents
for the three terms in the bound, otherwise one term will not go
to zero as the SNR goes to infinity. Fortunately, part of this has
already been accomplished in \cite{Zheng}.  Specifically, for the case where the
block length satisfies $T\ge M+N-1$ it was shown in \cite{Zheng} that
$d_{out}(r)=d_G(r)=d^*(r)$, although the $o(\log\mathrm{SNR})$ terms
are \em not \em the same. Hence, if we consider the asymptotic
regime of $\mathrm{SNR}$ approaching infinity we have

\begin{eqnarray*}
\lim_{\mathrm{SNR}\rightarrow\infty}\frac{\log
\bar{D}_{\tau}(\mathrm{SNR})}{\log
 \mathrm{SNR}}
 &\le&\lim_{\mathrm{SNR}\rightarrow\infty}\frac{\log\left[2^{-\frac{pTr}{k}\log\mathrm{SNR}+O(1)}+O(1)2^{-d^*(r)\log\mathrm{SNR}+o(\log\mathrm{SNR})}\right]}{\log\mathrm{SNR}}.\\
\end{eqnarray*}
If we choose an $r^*$ that solves
\begin{eqnarray}\label{equality}
d^*(r^*)&=&\frac{pTr^*}{k},
\end{eqnarray}
where $d^*(r)$ is the piecewise linear function connecting
$(N-r)(M-r)$ for integer values of $0<r<\min(M,N)$, then we have
\begin{eqnarray*}
\lim_{\mathrm{SNR}\rightarrow\infty}\frac{\log
\bar{D}_{\tau}(\mathrm{SNR})}{\log
 \mathrm{SNR}}&\le&\lim_{\mathrm{SNR}\rightarrow\infty}\frac{\log\left[2^{-d^*(r^*)\log\mathrm{SNR}+O(1)}+O(1)2^{-d^*(r^*)\log\mathrm{SNR}+o(\log\mathrm{SNR})}\right]}{\log\mathrm{SNR}}\\
              &\le&\lim_{\mathrm{SNR}\rightarrow\infty}\frac{\log\left[O(1)2^{-d^*(r^*)\log\mathrm{SNR}+o(\log\mathrm{SNR})}\right]}{\log\mathrm{SNR}}\\
              &=& -d^*(r^*).
\end{eqnarray*}

We now consider the lower bound (\ref{DTLB}) on average
distortion.  Again, for the
case where $T\ge M +N-1$ we have that $d_{out}(r)=d^*(r)$.  We can
match the exponents in (\ref{DTLB}) by choosing the same $r^*$
that satisfies (\ref{equality}), which yields
\begin{eqnarray*}
\lim_{\mathrm{SNR}\rightarrow\infty}\frac{\log
\bar{D}_{\tau}(\mathrm{SNR})}{\log\mathrm{SNR}}&\ge&
 \lim_{\mathrm{SNR}\rightarrow\infty}\frac{\log\left[2^{-\frac{pTr}{k}\log\mathrm{SNR}+O(1)}+O(1)2^{-d^*(r)\log\mathrm{SNR}+o(\log\mathrm{SNR})}\right]}{\log\mathrm{SNR}}\\
&\ge&
 \lim_{\mathrm{SNR}\rightarrow\infty}\frac{\log\left[2^{-d^*(r^*)+O(1)}+O(1)2^{-d^*(r^*)\log\mathrm{SNR}+o(\log\mathrm{SNR})}\right]}{\log\mathrm{SNR}}\\
&=& -d^*(r^*).
 \end{eqnarray*}
Since the asymptotic upper and lower bounds are tight, we have
proved the following theorem:

\bigskip
\noindent\textbf{Theorem 1:} In the limit of asymptotically high SNR, the optimal
end-to-end distortion for a vector quantizer cascaded with the
MIMO channel characterized by (\ref{channel}) satisfies
\begin{equation} d_D^* = 
 \lim_{\mathrm{SNR}\rightarrow\infty}\frac{\bar{D}_{\tau}(\mathrm{SNR})}{\log \mathrm{SNR}}=
-\min(d^*(r),pTr/k)=-d^*(r^*).
\end{equation}

The choice of optimal multiplexing rate $r^*$ is illustrated in Figure 2, which plots
$d^*(r)$ from Figure 1 together with $pTr/k$ as a function of $r$. We see
that the source distortion exponent $pTr/k$ {\em increases} linearly with $r$, while the 
 channel 
distortion exponent $d^*(r)$ {\em decreases}
 piecewise linearly with $r$. 
To balance the source and
channel distortion, $r^*$ is chosen such that
$d^*(r^*)=pTr^*/k$. 
\begin{figure}[h]
\begin{center}
\psfig{figure=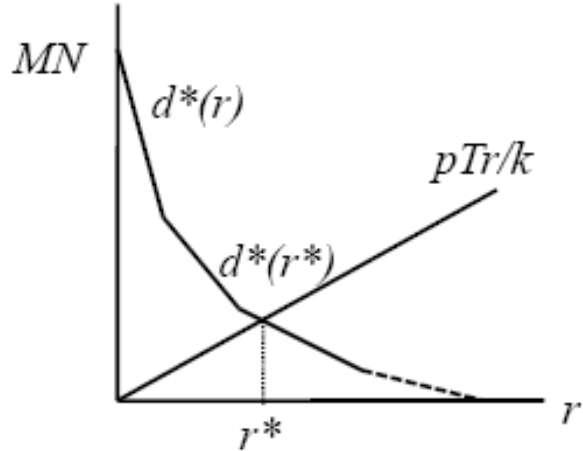, angle=0, width=3.5in}
\end{center}
\caption{ The optimal multiplexing rate $r^*$ to balance source and channel distortion.}
\end{figure}

\bigskip

\noindent It should be noted that the tightness of the above
bounds only hold when $T\ge M+N-1$.  For $T < M+N-1$ the upper
bound remains the same while the lower bound changes, which leaves a
gap between our bounds.

\subsection{Asymptotic Distortion Properties}

The asymptotic distortion and optimal distortion exponent from
Theorem 1 possess a few non-intuitive properties.  First, while it
is possible to choose  $d^*(r)=MN$ (full multiplexing) or $r=\min (M,N)$
(full diversity), it is never
optimal to do so.  When minimizing $\bar{D}_{\tau}(\mathrm{SNR})$ we require
non-zero amounts of both diversity and multiplexing, otherwise one
of the terms in the distortion bounds (\ref{DTUBave}) and (\ref{DTLB}) will not tend to zero as
$\mathrm{SNR}$ approaches infinity.  It is also interesting to
examine the optimal distortion exponent as the block length $T$ or
source dimension $k$ become large.  As $k$ becomes large (and $T$
remains fixed) we must increase $r^*$ in order to match the terms
in (\ref{equality}). This is consistent with our intuition since a high
dimensional source will require a large amount of multiplexing,
otherwise the distortion at the source encoder becomes very large.
It is more surprising that as $T$ becomes large (and $k$ remains fixed) we should decrease
$r^*$, i.e. increase diversity at the expense of multiplexing. This is 
in contrast to traditional source-channel
coding, where we encode our source at a rate just below the channel
capacity ($\min(M,N)\log\mathrm{SNR}$) when the block length
tends to infinity. In this case, however, we don't encode at channel capacity because
 the source dimension $k$
remains fixed as $T$ becomes large.  Thus, since the source encoding rate is
proportional to $T$, we are already getting an
asymptotically large channel rate for source
encoding, and therefore should use our antennas for diversity rather than additional rate
through multiplexing.

\subsection{Source-Channel Code Separation}

One feature that we do share with the traditional source-channel
coding results is the notion of separation.  In a traditional
Shannon-theoretic framework, the source encoder needs to know only
the channel capacity to design its source code.  Then one may
encode the source independently of the channel (at the channel
capacity rate) and achieve the optimal end-to-end distortion.  In
this case the end-to-end distortion is due only to the source
encoder since the channel is error free (over asymptotically long
block lengths).

In our model we consider a source encoder concatenated with a MIMO channel
that is restricted to transmission over finite block lengths.  
With this restriction the channel introduces errors even at transmission rates below capacity. These channel errors give rise to the diversity-multiplexing tradeoff. Under this finite blocklength channel coding we obtain a source and channel coding strategy to minimize end-to-end distortion. Our
results indicate that separate source and channel coding is still optimal for this
minimization. However, we now get (equal) distortion from both the source and channel code, in contrast to the optimal strategy in Shannon's separation theorem
where the source is encoded at a rate below channel capacity
and thus no distortion is introduced by the channel.

\subsection{Non-asymptotic Bounds}

We now analyze the behavior of our distortion bounds and the
corresponding choice of $r^*$ for finite
$\mathrm{SNR}$. In particular, we will consider the case of 
large but finite SNR, such that the SNR is sufficiently
large to neglect the $O(1)$ term in the exponent of
  (\ref{asym}) and (\ref{DTLB}), and
to assume $O(1) \approx 1$ and neglect the $o(\log SNR)$ exponential
term in (\ref{DTUBave}) and (\ref{DTLB}).
With these approximations the optimal diversity-multiplexing tradeoff is 
obtained by solving the following convex optimization problem:
\begin{eqnarray} \label{optimization}
\min_{r}&& 2^{-\frac{pT}{k}r\log \mathrm{SNR}}+2^{-d^*(r)\log \mathrm{SNR}}\\
\textrm{s.t.}&\;&0\le r\le\min(M,N).\nonumber
\end{eqnarray}

Figures 3, 4, and 5 provide numerical results based on the solution
to (\ref{optimization}) comparing the total end-to-end
distortion versus the number of antennas assigned to multiplexing.
Each plot contains four curves that represent different
$\mathrm{SNR}$ levels.  The difference between the three plots is
the ratio of the block length $T$ to source vector dimension $k$.
Notice that for $T$ much smaller than $k$ (Figure 3) we will use almost all
of our antennas for multiplexing. For $k$ of
the same order as $T$ (Figure 4) we will choose about the same number of
antennas for multiplexing and for diversity.
 For $k$ smaller than $T$ (Figure 5)
we will use more antennas for diversity than for multiplexing.  
Note that even at low SNR we can still find $r^*$ via the convex optimization
formulation in (\ref{optimization}), but
must include the neglected terms $O(1)$ and $o(\log SNR)$ in the distortion expressions
to which we apply this optimization. In our numerical results we found that
neglecting these terms for SNRs above 20 dB had little impact. 

\begin{figure}[h]
\begin{center}
\psfig{figure=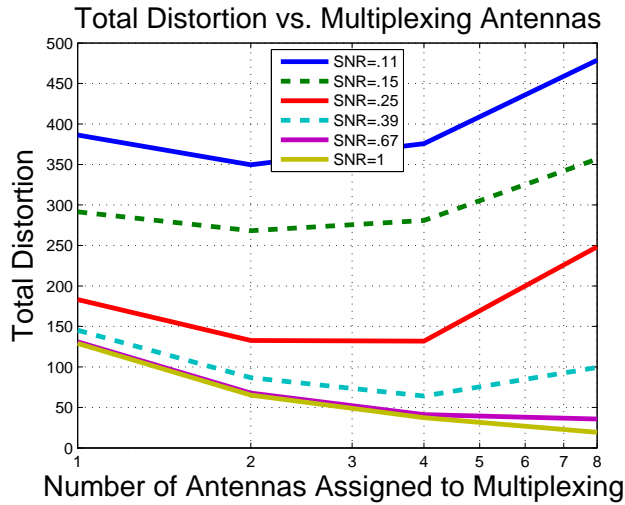, angle=0, width=3.5in}
\end{center}
\caption{ Total distortion vs. number of antennas assigned to multiplexing
in an 8x8 system ($T<<k$).} 
\end{figure}

\begin{figure}[h]
\begin{center}
\psfig{figure=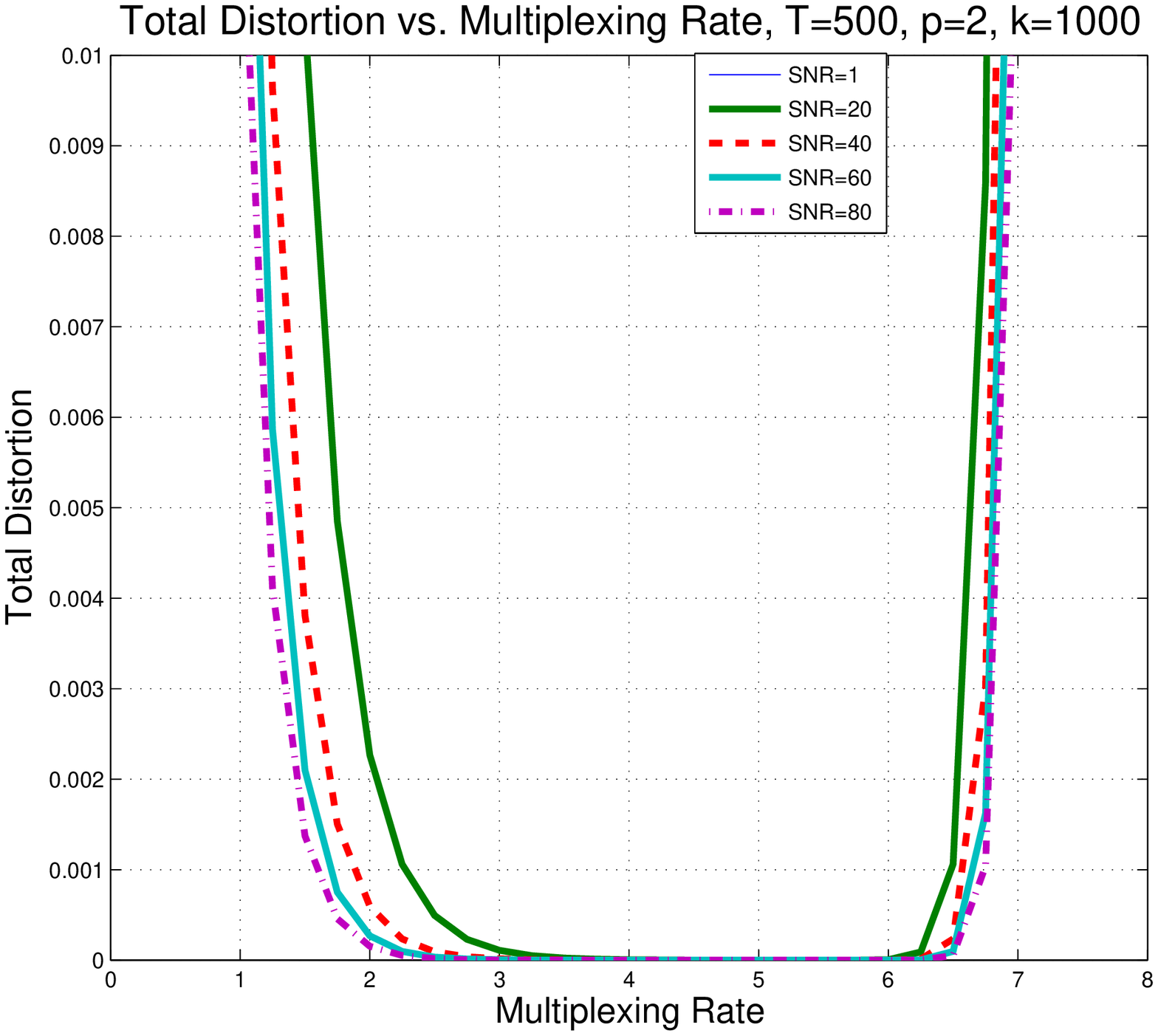, angle=0, width=3.5in}
\end{center}
\caption{ Total distortion vs. number of antennas assigned to
multiplexing in an 8x8 system ($T~k$).}
\end{figure}

\begin{figure}[h]
\begin{center}
\psfig{figure=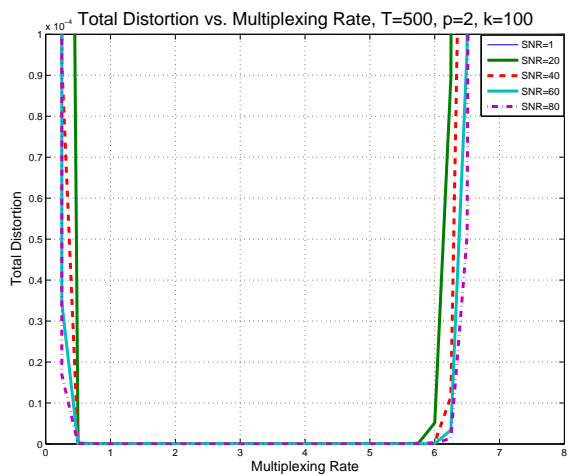, angle=0, width=3.5in}
\end{center}
\caption{ Total distortion vs. number of antennas assigned to
multiplexing in an 8x8 system ($T>>k$).} 
\end{figure}

\section{Practical Source and Channel Coding}

While the results in the previous section lead to closed form
solutions for optimal joint source-channel coding in the high SNR
regime, they only apply to a specific class of source and channel
codes and distortion metrics.  We now examine the
diversity-multiplexing tradeoff for a broad class of source codes,
channel codes, and distortion metrics.   The
basic optimization framework (\ref{optimization}) can still be applied to
this more general class of problems. Furthermore,
this framework can be applied in non-asymptotic settings,
thereby allowing us to study the diversity-multiplexing tradeoff
under typical operating conditions. In this section we present an example of
end-to-end distortion optimization, via the diversity-multiplexing
tradeoff, for source/channel distortion models that are fitted to
real video streams and MIMO channels.

We use the progressive video encoder model developed in
\cite{Girod}. The overall mean-square distortion is evaluated as
\begin{equation}
D_{\tau}=D_e+D_c,
\end{equation}
where $D_e$ is the distortion induced by the source encoder and
$D_c$ is the distortion created by errors in the channel. Although
the total distortion is represented by two separate components,
each component shares some common terms so we will still have a
tradeoff between diversity and multiplexing.
The model for source distortion $D_e$ developed in \cite{Girod}
consists of a six-parameter analytical formula
that is fitted to a particular traffic stream.  Numerical results for
$D_e$ as a function of the source encoding rate are provided
in \cite[Figure 2]{Girod}. The source encoder design is based
on a parameter $\beta$ corresponding to
the amount of redundant data in consecutive encoding
blocks.  In general a larger value of $\beta$ leads to a smaller $D_e$
at the cost of increased complexity.

The model for the channel distortion $D_c$ is fitted to the
following equation,
\begin{equation}
 D_c=\sigma^2 P_e(N_u)\left[ \frac{\gamma+\beta}{\gamma} \ln\left(1+\frac{\gamma}{\beta}\right)-\frac{1}{\gamma}+\frac{1}{2}
 \right],
\end{equation}
where given $\beta$ the parameters $\sigma^2$ and $\gamma$ are based
on the particular source encoder and traffic stream,
$N_u$ is the number of antennas used for multiplexing, and
$P_e(N_u)$ is the probability of codeword error as a function of
$N_u$. We will assume sources with $\beta=.01$ in our analysis since
it provides the lowest distortion for any given rate. This
source encoder setting also provides the highest sensitivity to
channel errors, which allows us to highlight the tradeoff between
multiplexing and diversity in our optimization.

Our channel transmission scheme follows the setup in \cite{Kuhn}.
We use 8 transmit and 8 receive antennas with a set of linear
space-time codes that can trade off multiplexing for diversity
(specifically, these codes only trade integer values of $r$ and
$(M,N)$). The actual code construction in \cite{Kuhn} is fairly
complex and involves several inner and outer codes designed to
handle both Ricean and Rayleigh fading channels in a MIMO
orthogonal frequency division multiplexing (OFDM) system. 
For the purposes of our numerical results
the actual code design is irrelevant, we only require
the probability of error as a function of SNR and the
number of antennas assigned to multiplexing,
which is given in \cite[Figure 4]{Kuhn}. 
Our optimization can be applied to space-time channel codes 
developed by  other authors
\cite{ElGamal,ElGamal2,Lu} by using the error probability associated
with their codes in our optimization. 

Since the channel coding scheme of \cite{Kuhn} does not permit us to assign
fractions of antennas, we must solve the following integer program
for the optimal distortion and number of multiplexing antennas:
\begin{eqnarray}
\min_{N_u}&& D_e+D_c\\
s.t. &\;& N_u\in\{1,2,4,8\}.\nonumber
\end{eqnarray}

Figure 6 contains a set of curves that show the total distortion
achieved as a function of the number of antennas assigned to
multiplexing.  The uppermost curve corresponds to the lowest
$\mathrm{SNR}$ and the bottom curve corresponds to the highest
$\mathrm{SNR}$. We see that we have an explicit tradeoff here that
depends on $\mathrm{SNR}$. At low $\mathrm{SNR}$ the total
distortion is minimized by assigning most antennas to diversity to
compensate for the high error probability in the channel.  As
$\mathrm{SNR}$ increases we assign more antennas to multiplexing
since this is a better use of antennas when the error probability
is low.  One significant difference between this plot and the
asymptotic results in Section IV is that here we do assign our
antennas to full multiplexing as the $\mathrm{SNR}$ becomes large.
The reason we observe this behavior is that the rate of our
codebook in this example does not scale with $\mathrm{SNR}$. 
Thus, as the
$\mathrm{SNR}$ becomes large we eventually reach a point where
distortion would be reduced by moving to a higher rate code that
is not available in the 8x8 space-time code under
consideration.  Hence, the optimal choice in this case is to
eventually move to full multiplexing. The implication of this
result is that  a MIMO system should have enough
 antennas to exploit full multiplexing at
 all available $\mathrm{SNR}$s.  A design framework for
such codes
has been developed in \cite{ElGamal2}, but the error probability analysis of these
codes is still needed to perform the joint
source-channel coding optimization.

\begin{figure}
\begin{center}
\psfig{figure=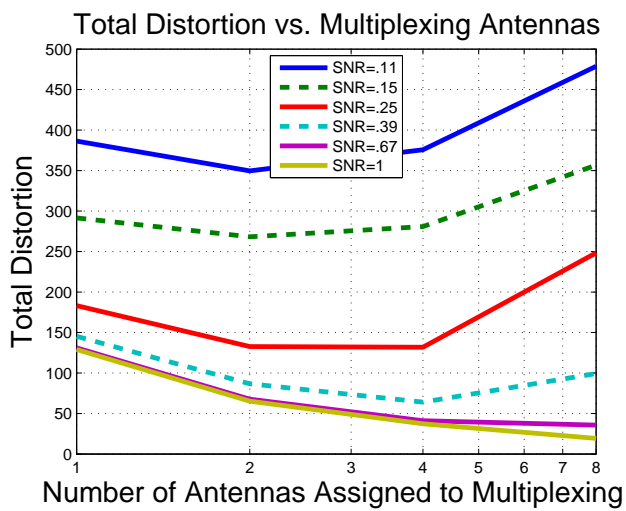, angle=0, width=3.5in}
\end{center}
\caption{ Total distortion vs. number of antennas assigned to
multiplexing for differing levels of SIR.}
\end{figure}

\section{The Diversity-Multiplexing-Delay Tradeoff}

Instead of accepting decoding errors in the channel, many 
wireless systems perform error correction via some form of ARQ.
In particular, the receiver has some form of error detection code,
and if a transmission error is detected on a given packet, a 
feedback path is used to send this error information back to the
transmitter, which then resends part or all of the packet to
increase the chance of successful decoding. 
The packet retransmissions, combined with random arrival times of
the messages at the
transmitter, cause queues to form in front of the source
coder and hence each block of data will experience random delays.
Here, the notion of delay we wish to capture is the time between
the arrival time of a message at the transmitter and 
the time at which it is successfully decoded at the receiver
(also known as the ``sojourn time'' in queueing systems).

While ARQ increases the probability of decoding a packet correctly,
it also introduces additional delay. The window size of the ARQ
protocol determines how many retransmission attempts will be made
for a given packet. The larger this window size, the more likely
the packet will be successfully received, and the larger the 
possible delays associated with retransmission will be. ARQ can
be viewed as a form of diversity, and hence it complements antenna
diversity in MIMO systems. For MIMO systems with ARQ, there
is a three-dimensional tradeoff between diversity due to multiple
antennas and ARQ, multiplexing, and delay. 
This three-dimensional tradeoff region was recently characterized
by El Gamal, Caire, and Damen
in \cite{ARQ}, and we will use this region in lieu of the 
Zheng/Tse diversity-multiplexing region in this section. 
We will first summarize results from \cite{ARQ} characterizing this
region, then use this region to optimize the
diversity-multiplexing-ARQ tradeoff for distortion under delay constraints.

\subsection{The ARQ Protocol and its Diversity Gain}

We assume the same $M$x$N$ channel model (\ref{channel}) as before and
the following ARQ scheme. 
 Each information message is encoded into a sequence of
$L$ blocks each of size $T$. Transmission commences with the first
block and after decoding the message the receiver sends a positive
(ACK) or negative (NACK) acknowledgement back to the transmitter.
In the case of a NACK the transmitter sends the next block in the
sequence and the receiver uses all accumulated blocks to try to
decode the message.  This process proceeds until either the
receiver correctly decodes the message or until all $L$ blocks
have been sent.  If a NACK is sent after the transmission of the
$L$th block then an error is declared, the message is removed from
the system, and the transmitter starts over with the next queued
message.  As in \cite{ARQ} we will use the term ``round''
to describe a single block transmission of length $T$.
We will refer to all $L$ rounds associated with the ARQ protocol as an
``ARQ block''. Hence, each ARQ block consists of up to $L$
rounds, and each round is of size $T$.

The fading coefficients $h_{ij}$ that model the gain from transmit
antenna $i$ to receive antenna $j$ are i.i.d. complex Gaussian
with unit variance.  The channel gain matrix $\mathbf{H}$ with
elements $H(i,j)=(h_{ij}:i\in\{1,\ldots N\},j\in\{1,\ldots,M\})$
is assumed to be known at the receiver and unknown at the
transmitter.  There are two channel models investigated in 
\cite{ARQ}: the long-term static model and the short-term static model.
In the long-term static model  the channel remains constant over each
ARQ block of up to $LT$ symbols, and the fading associated with each ARQ block is
i.i.d. In the short-term static model the fading is constant over one ARQ
round, then changes to a new i.i.d. value. The long-term model applies to a
quasi-static situation such as might be seen in a wireless LAN channel. The
short-term model is more dynamic and might correspond to fading associated with
a portable mobile device. The ARQ diversity gain is very
similar for the two models. In particular, the diversity exponent for the short-term
static model is a factor of $L$ larger than for the long-term static model,
corresponding to the $L$-fold time diversity in the short-term model. We will
use the long-term static model in our analysis and numerical results, since
it allows us to focus on the diversity associated with the ARQ rather than time
diversity. Our analysis easily extends to the short-term static model by adding the 
extra factor of $L$ to the ARQ diversity exponent. 

Under the long-term static channel model, in round $l\in\{1,\ldots,L\}$ of an ARQ block we
can represent the channel as
\begin{equation}
\mathbf{Y_l}=\sqrt{\frac{\mbox{SNR}}{M}}\mathbf{HX_l}+\mathbf{W},
\end{equation}
where $\mathbf{X_l}\in\mathcal{C}^{M\mathrm{x}T}$ and
$\mathbf{Y_l}\in\mathcal{C}^{N\mathrm{x}T}$ are the transmitted
and received signals in block $l$, respectively.  The additive
noise vector $\mathbf{W}$ is i.i.d. complex Gaussian with unit
variance.

With the above model in hand let us define a family of codes
$\{C(\mbox{SNR})\}$, indexed by the $\mbox{SNR}$ level.  Each code
has length $LT$ and the bit rate of the first block in each code
is $b(\mbox{SNR})/T$. Suppose we consider a sequence of ARQ
blocks. At time $s$ the random variable $B[s]=b(\mbox{SNR})$ if a
message is successfully decoded at the receiver, and $B[s]=0$
otherwise. Then, we can define the average throughput of the ARQ
protocol using these codes as
\begin{equation}
\eta(SNR)=\liminf_{\tau\rightarrow\infty}\frac{1}{T\tau}\sum_{s=1}^{\tau}B[s],
\end{equation}
and we can view $\eta(\mbox{SNR})$ as the average number of
transmitted bits per channel use.  Further define
$P_e(\mbox{SNR})$ as the average probability of error of the ARQ
block (i.e. the probability that a NACK is sent after $L$
transmission rounds). The multiplexing gain of the ARQ
protocol is defined in \cite{ARQ} as
\begin{equation}
r=\lim_{\mbox{SNR}\rightarrow\infty}\frac{\eta(\mbox{SNR})}{\log
\mbox{SNR}},
\end{equation}
and the diversity gain as
\begin{equation}
d=-\lim_{\mbox{SNR}\rightarrow\infty}\frac{\log
P_e(\mbox{SNR})}{\log \mbox{SNR}}.
\end{equation}

For each $r$ and $L$ we define the optimal diversity gain
$d^*(r,L)$ as the supremum of the diversity gain achieved by any
scheme.  For $L=1$ (i.e. no ARQ) we have the original
diversity-multiplexing tradeoff from Section II.  Hence, $d^*(r,1)$
is the piecewise linear function $d^*(r)$ joining the points
$(k,(M-k)(N-k)$, at integer values of $k$ for $0\le k\le
\min(M,N)$.  For $L>1$ we have the following result from
\cite{ARQ}.

\bigskip
\noindent\textbf{Diversity Gain of ARQ:}  The diversity gain for
the ARQ protocol with a maximum of $L$ blocks is
\begin{equation}\label{dude}
d^*(r,L)=d^*\left(\frac{r}{L}\right).
\end{equation}
\bigskip

The diversity gain achieved by ARQ is quite remarkable. According
to (\ref{dude}), for any $r<\min(M,N)$ we can achieve the full
diversity gain $d=MN$ for sufficiently large $L$.  Thus, for $L$ sufficiently
large, there is
no reason to utilize spatial diversity since all needed diversity can
be obtained through ARQ. For $L$ not sufficiently large, the maximum ARQ window
size would still be utilized to minimize the amount of spatial diversity required.
 The diversity-multiplexing-ARQ tradeoff (\ref{dude}) is analogous to the Zheng-Tse
diversity-multiplexing 
tradeoff $d^*(r)$. Thus,  the same
analysis as in Section III can be applied to minimize end-to-end distortion based on
the diversity-multiplexing tradeoff $d^*(r,L)$ induced by the ARQ. In 
particular, end-to-end distortion 
for MIMO channels with asymptotically high SNR
and ARQ retransmissions, in the absence of a delay constraint, is
minimized
using the following procedure:
\begin{enumerate}
\item choose the largest ARQ window size $L$ possible,
 \item determine the resulting ARQ diversity gain $d^*(r,L)$ from (\ref{dude})
 \item solve (\ref{equality}) for the optimal rate $r^*$ using $d^*(r,L)$ instead of $d^*(r)$.
 \end{enumerate}
This procedure not only minimizes end-to-end distortion, 
but also indicates that separate source and channel coding is optimal, 
provided the source and channel encoders know $r^*$ and
the maximum value of $L$.  Moreover, the results in \cite{ElGamal}
show that the rate penalty for ARQ is negligible in the high SNR
regime.

In order to analyze the diversity, multiplexing, and delay tradeoff
for delay-sensitive sources we must recognize two important
subtleties about the above results.  First, in systems that transmit
delay-constrained traffic we may not be able to tolerate a long
ARQ window (in some cases ARQ may not be tolerated at all).
Second, we must carefully consider the impact of asymptotically high SNR,
which is crucial in the proofs of the above
results. Specifically, in the high SNR regime the occurrence of a
NACK in the ARQ protocol becomes a rare event (i.e. the
probability of a NACK tends to zero as $\mbox{SNR}$ approaches
infinity). Therefore, with probability tending to one, each
message is decoded correctly during the first transmission attempt
-- resulting in a multiplexing gain equivalent to that of a system
without ARQ.  The increasingly rare errors are corrected by the
ARQ process, which results in increased diversity.

The main difficulty in using these asymptotic results to evaluate
delay performance is that in the high SNR regime there is
essentially \em no delay \em due to ARQ.  In other words, queuing 
delays associated with  retransmissions are rare in the high SNR
regime.  Based on this fact and using standard results from queuing theory, one can show
that under stable arrival rates the arriving messages almost 
always find the system empty. Hence,
with high probability an arriving message will immediately begin
transmission and suffer no queuing delay. In wireless systems,
errors during a transmission attempt are not rare events. Indeed,
most wireless systems typically become reliable only after the
application of ARQ.  In other words, errors after completion of
the ARQ process might be rare events, but errors \em during \em
the ARQ process are not rare. As we shall see in the next
subsection, this subtle difference requires a 
an optimization framework that can model and optimize over the queuing dynamics
associated with ARQ.

\subsection{Delay-Distortion Model} 
This section presents our model for a  delay-sensitive system. 
  We do not assume a high
SNR regime in our analysis since, as stated in the previous section,
this leads to rare ARQ errors and hence effectively removes the
ARQ queuing delay.  We do assume that the finite SNR
is fixed for each problem instance, i.e. we do not
optimize power control, although this optimization was investigated in
\cite{ARQ} and shown to provide significant diversity gains in the
long-term static channel.

We assume the original source data $u$ is a random vector with
probability density $h(u)$, which has support on a closed, bounded
subset of $\Re^k$ with non-empty interior.  During each
transmission block of length $T$ an instance of $u$ arrives at the
system independently with probability $\lambda$ and is queued for
transmission. We assume that each message has a deadline
$k$ at the receiver. Hence, if a message arrives at time $t$ and
is not received by time $t+kT$ then its deadline expires and the
message is dropped from the system. We assume that each message is
quantized according to the scheme discussed below.  The quantized
version of each message is then mapped into a codeword in the
codebook $\{C(\mbox{SNR})\}$ and passed to the MIMO-ARQ
transmitter discussed in the previous section.

Due to the random message arrival times and the random completion
times of the ARQ process we will have queuing and delay in this
system.  Our goal is to select a diversity gain, multiplexing
gain, and ARQ window size to minimize the distortion created by
both the quantizer and the messages lost due to channel error or delay.
 The intuition behind the diversity-multiplexing-ARQ tradeoff is
straightforward. We would like to use as much multiplexing as
possible since this will allow us to use more bits to describe a
message and reduce encoder distortion. However, high levels of
multiplexing induce more errors in the wireless channel, thereby
requiring longer ARQ windows to reduce errors.  The longer ARQ
windows induce higher delays, which also cause higher distortion
due to messages missing their deadlines.  We must balance all of
these quantities to optimize system performance.

We use the same vector encoder and distortion model from Section
III. As before, we assume that the total average distortion
$D_{\tau}(F,\mbox{SNR})$ can be split into two dependent pieces
\begin{equation}
D_{\tau}(F,SNR)= D_s(F) + D_e(d,\mbox{SNR}),
\end{equation}
where $D_e(d,\mbox{SNR})$ is the distortion caused by messages
declared in error. Here the errors are incurred whenever the ARQ
process fails or when a message's deadline expires.  We also
assume the distortion due to erroneous messages is bounded by the
overall loss probability:
\begin{equation}\label{dude1}
D_e(d,\mbox{SNR})\le P_e(\mbox{SNR})+P\{Delay>k\},
\end{equation}
where $P\{Delay>k\}$ is the probability that a message violates
its deadline and $P_e(\mbox{SNR})$ is the probability of
error for the ARQ block, which depends on its window size $L$. 

Our goal is to minimize the total delay-distortion bound
\begin{eqnarray}\label{dude3}
D_{\tau}(F,\mbox{SNR})\le D_s(F)+P_e(\mbox{SNR})+P\{Delay>k\}.
\end{eqnarray}
In order to optimize (\ref{dude3}) we require a formulation that
accounts for the different delays experienced by each message.
Hence, as described in the next section,
 we turn to the theory of Markov decision processes to model
and solve this problem.

\subsection{Minimizing Distortion via Dynamic Programming}

We now develop a dynamic programming optimization framework to 
minimize (\ref{dude3}). 
We assume without loss of generality that the queue 
in our system is of maximum size $k$.  This is not a restrictive
assumption since each message
requires at least one time block of size $T$ for transmission,
hence any arriving message that sees more than $k$ messages in the
queue will not be able to meet its deadline and could be dropped
without affecting our performance analysis.  
Note that unlike standard
queuing models that only track the number of messages awaiting
transmission, we must also track the amount of time a particular
message has waited in the queue. In particular, given that one message is queued for
transmission our state space model must differentiate between a
message that has just arrived and a message whose deadline is about to
expire.  Since the queue size is bounded, 
we can only have a finite number of messages in the
queue, and hence the combined message and waiting time model exists in a
finite space.  

We define the queue process $X_Q=(X_Q(n):n\ge 0)$,
which takes values on a finite space $\mathcal{X_Q}$.  Similarly, we
define the state of the ARQ process $X_L=(X_L(n):n\ge 0)$
on a finite space $\mathcal{X_L}$.  Here, the state of the ARQ
process denotes the number of the current transmission round in
the current ARQ block. Finally, we define the overall state of the system as a
process $X=(X(n):n\ge 0)$ such that $X(n)=(X_Q(n),X_L(n))$ (i.e.
the space $\mathcal{X}$ is the product space of $\mathcal{X_Q}$
and $\mathcal{X_L}$).

Since the arrival process is geometric and each ARQ round is
assumed to be i.i.d., the process $X$ is a finite-state
discrete-time Markov chain.  The transition dynamics of this
Markov chain are governed by the choices of diversity,
multiplexing, and the ARQ window size.  We assume that at the start of
each ARQ block the transmitter chooses the number of bits to
assign to the vector encoder and hence the amount of spatial
diversity and multiplexing in the codeword selected from
$\{C(\mbox{SNR})\}$.  The transmitter also selects the length of
the ARQ window.  These choices then remain fixed until either the
message is received or the ARQ window expires.  Define the space
of actions $\mathcal{A}$ as the set of all possible combinations
of multiplexing gain and ARQ window length.  Note that a choice
of multiplexing gain implicitly selects the number of bits given
to the source encoder as well as the amount of spatial diversity.  
We assume that the number of antennas $M$ and $N$ are
finite and that the ARQ window size is also finite.  Hence, the
action space $\mathcal{A}$ is a finite set.

We define the control policy $g$ as a probability distribution on the
space $\mathcal{X} \; \mathrm{x} \; \mathcal{A}$.  We can view the
elements of $g$ as
\begin{eqnarray*}
g(x,a)=P\{\mathrm{action} \; a \;
\mathrm{chosen}\;\mathrm{in}\;\mathrm{state}\;x\},\; \forall
x\in\mathcal{X}\; ,a\in\mathcal{A}.
\end{eqnarray*}
For any control $g$, the Markov chain $X$ is irreducible and
aperiodic\footnote{To create a non-irreducible Markov chain we
would be required to successfully transmit a packet with
probability one.}. Define $Q(g)$ as the transition matrix for $X$
corresponding to control policy $g$. Hence, $Q(g)=(Q_{i,j}(g):i,j
\in \mathcal{X})$ is a stochastic matrix with entries
\begin{eqnarray*}
 \lefteqn{Q_{i,j}(g)=P\left(X(n+1)=j|X(n)=i,g\right)}\\&=&\sum_{a\in\mathcal{A}}P\left(X(n+1)=j|X(n)=i,A(n)=a)g(i,a)\right).
\end{eqnarray*}
For each state-action pair we define a reward function $r(x,a)$.
For the states in $\mathcal{X}$ corresponding to completion of the
ARQ process the reward function denotes the distortion incurred in
that particular state.  Hence,
\begin{equation}\label{reward}\nonumber
 r(x,a)=2^{-ps/k}+I[\mathrm{ARQ}\;\;\mathrm{Fails}]+I[Delay>k].
\end{equation}

 Let $\mathcal{G}$ be the set of all available control policies.
 Then for any $g\in\mathcal{G}$ define the limiting average value of $g$ starting from state $x$ as
\begin{eqnarray*}
 V(x,g)&=& \limsup_{n\to\infty}  \left [ \left ( \frac{1}{n+1}\right) \sum_{k=0}^{n} E_{x,g} \left[r(X(k),g) \right ] \right
 ],
\end{eqnarray*}
where $r(X(k),g)$ is the random reward earned at time $k$ under
control policy $g$.  Since $X$ is an irreducible and aperiodic
Markov chain for any control $g$ we know from \cite{Bertsekas} that the
above value function reduces to
\begin{equation}
V(x,g)=\pi(g)r(g) \; \forall \;x\in\mathcal{X},
\end{equation}
where $\pi(g)=\pi(g)Q(g)$ is the stationary distribution of $X$
under control $g$ and $r(g)$ is the column vector of rewards
earned for each state $x\in\mathcal{X}$ under control $g$.  Hence,
the value function is simply the expected value of our reward
function $r$ with respect to the stationary distribution of $X$.
Notice that given our definition for $r$ in (\ref{reward}), the
value function $V(g)$ provides us with the delay-based distortion
(\ref{dude3}) caused by control policy $g$. Thus we want to minimize
distortion by minimizing the value function $V(g)$. 

Specifically, our goal is to find a $g \in \mathcal{G}$ that minimizes $V(x,g)$.
From \cite{Bertsekas} we know this problem can be solved through
the following linear program.
\begin{equation} \label{linprog}
\min_{s} \sum_{x \in \mathcal{X}} \sum_{a \in \mathcal{A}}
r(x,a)s_{x a}
\end{equation}
subject to:
\begin{eqnarray*}
\sum_{x \in \mathcal{X}} \sum_{a \in \mathcal{A}} \left (
\delta(x,x\, ')-p(x\, '|x,a) \right)s_{x a} &=& 0, \; x\, ' \in
\mathcal{X}
\label{constraint} \nonumber \\
\sum_{x \in \mathcal{X}} \sum_{a \in \mathcal{A}} s_{x a} &=& 1, \nonumber \\
s_{x a} \ge 0; \; a \in \mathcal{A}, \; x \in
\mathcal{X}, \nonumber \\
\end{eqnarray*}
where $\delta(x,x\, ')$ is the Kronecker delta, $s_{x a}$ is the
steady-state probability of being in state $x$ and taking action
$a$, and $p(x\, '|x,a)$ is the probability of jumping to state
$x\, '$ given action $a$ in state $x$. The state-action
frequencies $s_{x a}$ provide a unique mapping to an optimal
control ${g}^{*}$ \cite{Bertsekas}.

With this dynamic programming formulation in hand we can solve for
the optimal diversity gain, multiplexing gain, and ARQ window size
as a function of queue state and deadline sensitivity.  We
demonstrate the performance of these solutions with a numerical
example in the next subsection.

\subsection{Distortion Results}

Consider the ARQ system described above with messages arriving
in each time block with probability $\lambda=0.9$.  We assume a
4x4 MIMO-ARQ system ($M=N=4$) with an SNR of 10 dB
that utilizes the incremental redundancy
codes proposed in \cite{ElGamal2}, which have been shown to achieve
the diversity-multiplexing-ARQ tradeoff.  For these codes
we allow the ARQ window size to
take values in a finite set $L \in \{1,\ldots,4\}$. We also consider
the deadline length $k$ ranging over several values
($k\in\{2,\ldots,8\}$) to examine the impact of delay
sensitivity on the solution to our dynamic program
(\ref{linprog}).  For each value of $k$ we solve a new version of
(\ref{linprog}). The plots below contain the data accumulated by
averaging over 
all of these solutions.

Figure 7 plots the optimal ARQ window length as a function of
queue state for different values of $k$.  We see that for short
deadlines we cannot afford long ARQ windows for any queue state.
As the deadlines become more relaxed we can increase the ARQ
window size.  However as the queue fills up we are forced to again
decrease the amount of ARQ diversity.

\begin{figure}[h]
\begin{center}
\psfig{figure=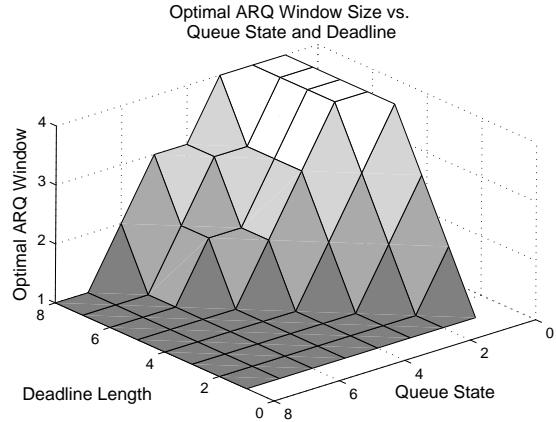, angle=0, width=3.25in}
\end{center}
\caption{Optimal ARQ window size vs. queue state vs. deadline
length $k$ (SNR=10 dB).}
\end{figure}

Figure 8 plots the optimal multiplexing gain $r$ as a function of
queue state for different values of $k$.  Here we see that with
short deadlines we must use fairly low amounts of spatial
multiplexing (i.e. high spatial diversity), since we cannot use
ARQ diversity.  As the deadlines become more relaxed we can
increase the amount of spatial multiplexing and use ARQ for
diversity.  Once again, as the queue fills up we must switch back
to low levels of multiplexing or, equivalently, high levels of diversity
 to ensure a lower error probability and hence that fewer retransmissions are 
needed to clear a given message from the system.

\begin{figure}[h]
\begin{center}
\psfig{figure=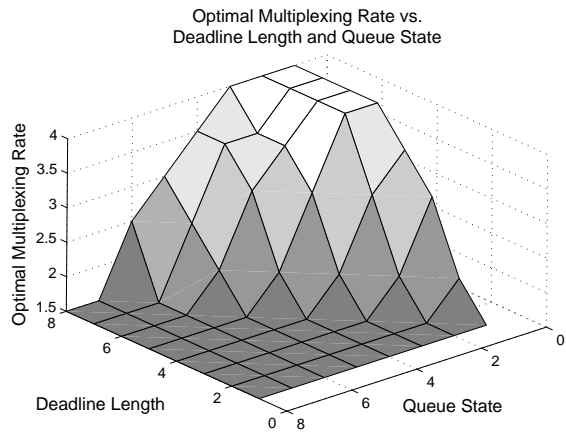, angle=0, width=3.25in}
\end{center}
\caption{Optimal multiplexing gain vs. queue state vs. deadline
length $k$ (SNR=10 dB).}
\end{figure}

We also evaluate the performance advantage gained by adapting the
settings of diversity, multiplexing, and ARQ rather than choosing
fixed allocations.  For $k=4$ we computed the distortion resulting
from all possible fixed allocations of ARQ window length and
multiplexing gain. The curved surface in Figure 9 plots the
distortion of these fixed allocations for all values of $L$ and
$r$.  The flat surface in Figure 9 is the distortion achieved by
the adaptive scheme (plotted as a reference), which indicates a 
distortion reduction of up to 70 dB.  Even in the most
favorable cases, the adaptive scheme outperforms any fixed scheme
by more than 50\%. 

\begin{figure}[h]
\begin{center}
\psfig{figure=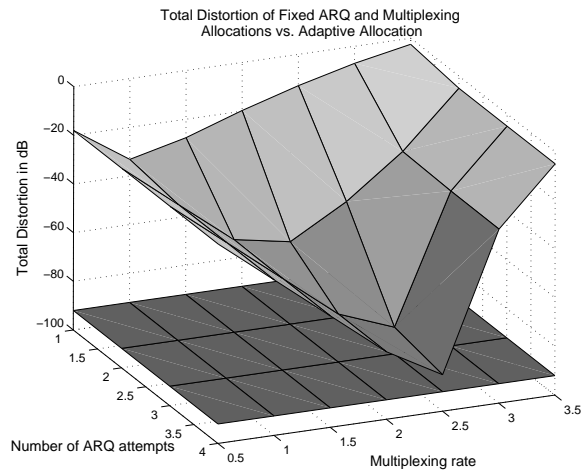, angle=0, width=3.25in}
\end{center}
\caption{Distortion for the fixed allocation problem vs.
multiplexing gain vs. ARQ window size (SNR=10 dB).}
\end{figure}

\section{Summary}

We have investigated the optimal tradeoff between
diversity, multiplexing, and delay in MIMO systems to 
minimize end-to-end distortion under both asymptotic assumptions
as well as in practical operating conditions. We first considered
the tradeoff between diversity and multiplexing without a delay constraint.
In particular, for the asymptotic
regime of high SNR and source dimension, we obtained a 
closed-form expression for the optimal rate on the 
Zheng/Tse diversity-multiplexing tradeoff region as a simple function
of the source dimension, code blocklength, and distortion norm. 
We also showed that in this asymptotic regime separate source and channel coding at the
optimized rate 
minimizes end-to-end distortion. However, in contrast to codes designed according to
Shannon's separation theorem, the finite blocklength assumption in our setting
causes distortion to be introduced
by both the source code and the channel code, even though the source encoding
rate is below channel capacity. We showed that the same optimization framework
can be applied even without an asymptotically large SNR. However, outside this asymptotic
regime, closed-form expressions for the optimal diversity-multiplexing tradeoff (and
corresponding transmission rate) cannot be found, and convex optimization tools are
required to find this optimal operating point. Finally, we developed an optimization framework to minimize end-to-end
distortion for a broad class of practical source and channel codes, and
applied this framework to a specific example of a video source code and space-time channel
code. Our numerical results illustrate quantitatively how the optimal number of antennas used
for multiplexing increases with both the source rate and the SNR.

We then extended our analysis to delay-constrained sources and MIMO systems using an
ARQ retransmission protocol. ARQ provides additional diversity in the system at the expense of 
delay. Minimizing end-to-end delay thus entails
finding the optimal operating point on the 
diversity-multiplexing-delay
tradeoff region. 
We developed a dynamic programming formulation for this optimization to capture the 
diversity-multiplexing tradeoffs of the channel as well as the dynamics of random message arrival times and random ARQ block completion times. The dynamic program
can be solved using standard techniques, which we applied to a 4x4 MIMO system with 
different 
ARQ window sizes and delay constraints. 
We obtained numerical results indicating the optimal amount of diversity, multiplexing,
and ARQ to use as a function of the queue state and message deadline. We also
demonstrated that adaptation of the diversity-multiplexing characteristics of the
MIMO channel code to the time-varying
backlog in the system leads to distortion reduction of up to 70 dB versus a
static allocation. 

The unconsummated union between information theory and networks has
vexed both communities for many years. As pointed out in \cite{Ephremides},
part of the reason for this disconnect is that source burstiness and
end-to-end delay are major components in the study of networks, yet
play little role in traditional Shannon theory where 
delay is asymptotically infinite and channel capacity inherently assumes a source with
infinite data to send. We hope that our work provides one small
step towards consummating this union by merging
information-theoretic tradeoffs associated with the channel with
models and analysis tools from networking to handle source burstiness and system delay.  Much
work remains to be done in this area by extending our ideas and developing
new ones for coupling the fundamental performance limits of general multihop networks
with queuing delay, traffic statistics, and end-to-end metric optimization
for heterogeneous applications running over these networks.

\section{Acknowledgments}
We are deeply grateful to the four reviewers for their detailed and
insightful comments, which helped to greatly improve the clarity and exposition of
the paper. We want to thank Reviewer D in particular for suggesting
Figure 2 to illustrate the optimization of the multiplexing rate $r^*$.

\newpage

\begin{center} {\bf List of Figures and Captions} \end{center}
\begin{itemize}
  \item Figure 1: The optimal diversity-multiplexing tradeoff for $T\ge
M+N-1$
  \item Figure 2: The optimal multiplexing rate $r^*$ to balance source and channel distortion
  \item Figure 3: Total distortion vs. number of antennas assigned to multiplexing
in an 8x8 system ($T<<k$)
  \item Figure 4: Total distortion vs. number of antennas assigned to
multiplexing in an 8x8 system ($T~k$)
  \item Figure 5: Total distortion vs. number of antennas assigned to
multiplexing in an 8x8 system ($T>>k$)
  \item Figure 6: Total distortion vs. number of antennas assigned to
multiplexing for differing levels of SIR.
  \item Figure 7: Optimal ARQ window size vs. queue state vs. deadline
length $k$ (SNR=10 dB)
  \item Figure 8: Optimal multiplexing gain vs. queue state vs. deadline
length $k$ (SNR=10 dB)
  \item Figure 9: Distortion for the fixed allocation problem vs.
multiplexing gain vs. ARQ window size (SNR=10 dB)
 \end{itemize}

\pagebreak
 \begin{center} {\bf Author Biographies} \end{center}
{\bf Holliday:} Tim Holliday received the B.S. degree in general engineering
from Harvey Mudd College, Claremont, CA, in 1997; the M.S. degree in electrical
engineering from Stanford University, Stanford, CA, in 2001; and the
Ph.D. degree in management science and engineering from Stanford University
in 2004.
His industry experience includes summer internships at Lucent Technologies
Bell Labs in summers of 2000 and 2001. He also served as a Communications
Officer in the U.S. Air Force Reserve from 1997 through 2004. From 2004-2006 he was
a Postdoctoral Research Associate at Princeton University, Princeton, NJ. 
In 2007 he joined Goldman Sachs as an associate. His
research interests include stochastic processes and modeling, cross-layer design
in wireless communications, and information theory.

\vspace{.3in}
\noindent {\bf Goldsmith:} Andrea J. Goldsmith is a professor of Electrical Engineering at Stanford University, and
was previously an assistant professor of Electrical Engineering at Caltech. She has also
held industry positions at Maxim Technologies and at AT\&T Bell Laboratories, and is
currently on leave from Stanford as co-founder and CTO of  Quantenna Communications,
Inc. Her research includes work on capacity of wireless channels and networks, wireless
communication and information theory, energy-constrained wireless communications,
wireless communications for distributed control, and cross-layer design of wireless
networks. She is author of the book ``Wireless Communications'' and co-author of the
book ``MIMO Wireless Communications,'' both published by Cambridge University
Press. She received the B.S., M.S. and Ph.D. degrees in Electrical Engineering from U.C.
Berkeley.

Dr. Goldsmith is a Fellow of the IEEE and of Stanford. She has received several awards
for her research, including the National Academy of Engineering Gilbreth Lectureship,
the Alfred P. Sloan Fellowship, the Stanford Terman Fellowship, the National Science
Foundation CAREER Development Award, and the Office of Naval Research Young
Investigator Award. She was also a co-recipient of the 2005 IEEE Communications
Society and Information Theory Society joint paper award. She currently serves as
associate editor for the IEEE Transactions on Information Theory and as editor for the
Journal on Foundations and Trends in Communications and Information Theory and in
Networks. She was previously an editor for the IEEE Transactions on Communications
and for the IEEE Wireless Communications Magazine, and has served as guest
editor for several IEEE journal and magazine special issues. Dr. Goldsmith is active in
committees and conference organization for the IEEE Information Theory and
Communications Societies and is an elected member of the Board of Governors for
both societies. She is a distinguished lecturer for the IEEE Communications Society, the
vice-president and student committee founder of the IEEE Information Theory Society, and was the technical program co-chair for the 2007 IEEE International Symposium on 
Information Theory. 

\vspace{.3in}
\noindent {\bf Poor:} H. Vincent Poor (S'72, M'77, SM'82, F'87) received the Ph.D. degree in EECS from Princeton University in 1977.  From 1977 until 1990, he was on the faculty of the University of Illinois at Urbana-Champaign. Since 1990 he has been on the faculty at Princeton, where he is the Dean of Engineering and Applied Science, and the Michael Henry Strater University Professor of Electrical Engineering. Dr. Poor's research interests are in the areas of stochastic analysis, statistical signal processing and their applications in wireless networks and related fields. Among his publications in these areas are the recent book MIMO Wireless Communications (Cambridge University Press, 2007), co-authored with Ezio Biglieri, et al, and the forthcoming book Quickest Detection (Cambridge University Press, 2008), co-authored with Olympia Hadjiliadis.

Dr. Poor is a member of the National Academy of Engineering, a Fellow of the American Academy of Arts and Sciences, and a former Guggenheim Fellow. He is also a Fellow of the Institute of Mathematical Statistics, the Optical Society of America, and other organizations.  In 1990, he served as President of the IEEE Information Theory Society, and in 2004-07 as the Editor-in-Chief of these Transactions. Recent recognition of his work includes the 2005 IEEE Education Medal, the 2007 IEEE Marconi Prize Paper Award, and the 2007 Technical Achievement Award of the IEEE Signal Processing Society.

\end{document}